\documentclass[conference]{IEEEtran}
\IEEEoverridecommandlockouts
\usepackage{cite}
\usepackage{amsmath,amssymb,amsfonts}
\usepackage{algorithmic}
\usepackage{graphicx}
\usepackage{textcomp}
\usepackage{xcolor}
\usepackage{booktabs} 
\usepackage{braket}
\usepackage{hyperref} 
\def\BibTeX{{\rm B\kern-.05em{\sc i\kern-.025em b}\kern-.08em
    T\kern-.1667em\lower.7ex\hbox{E}\kern-.125emX}}
\begin{document}

\title{Scalable Tensor Network Simulation for Quantum–Classical Dual Kernel

\thanks{\IEEEauthorrefmark{2} Corresponding Author:2503001@niar.org.tw}
}

\author{
\IEEEauthorblockN{
    Mei Ian Sam \IEEEauthorrefmark{1} and 
    Tai-Yu Li \IEEEauthorrefmark{2} 
}
\IEEEauthorblockA{\IEEEauthorrefmark{1}Department of Physics, National Tsing Hua University, Hsinchu, Taiwan}
\IEEEauthorblockA{\IEEEauthorrefmark{2}National Center for High-performance Computing, National Institutes of Applied Research, Hsinchu, Taiwan}
}


\maketitle

\begin{abstract}
This paper presents an efficient and scalable tensor network framework for quantum kernel circuit simulation, alleviating practical costs associated with increasing qubit counts and data size. The framework enables systematic large-scale evaluation of a linearly mixed quantum–classical dual kernel of up to 784 qubits. Using Fashion-MNIST, the classification performance of the test dataset is compared between a classical kernel, a quantum kernel, and the quantum–classical dual kernel across the feature dimensions from 2 to 784, with a one-to-one mapping between encoded features and qubits. Our result shows that the quantum–classical dual kernel consistently outperforms both single-kernel baselines, remains stable as the dimensionality increases, and mitigates the large-scale degradation observed in the quantum kernel. Analysis of the learned mixing weights indicates that quantum contributions dominate below 128 features, while classical contributions become increasingly important beyond 128, suggesting that the classical kernel provides a stabilizing anchor against concentration effects and hardware noise while preserving quantum gains at lower dimensions.
\end{abstract}


\section{Introduction}
In recent years, the co-evolution of deep learning and quantum hardware has pushed quantum machine learning (QML) to the forefront of interdisciplinary research, motivating a broad family of near-term learning paradigms designed to remain practical on noisy intermediate-scale quantum devices (NISQ) \cite{Biamonte2017QML,Preskill2018NISQ}. Representative directions include variational quantum algorithms  \cite{Cerezo2021VQA}, quantum neural-network-style architectures \cite{Killoran2019CVQNN,anQPW2025quantum,anQGAT2025,yuchaoQAE2025,chonweiQIR2025}, sequence models such as quantum-LSTM \cite{Chen2020QLSTM,hsuQKlstm2025quantum,hsuQKlstmAQI2025quantum}, expressivity-enhancing data re-uploading \cite{PerezSalinas2020Reuploading}, and quantum generative models such as qGANs \cite{Zoufal2019qGAN} and chemical molecular generation \cite{lyQMG2025}. Across these approaches, a recurring theme is to encode data using quantum feature maps into high-dimensional Hilbert spaces and perform a learning or similarity evaluation in that space under NISQ constraints \cite{Preskill2018NISQ}.

Within this landscape, quantum kernel methods are often viewed as a comparatively robust near-term route because they interface directly with mature classical kernel learners \cite{SchuldKilloran2019FeatureHilbert}. A quantum feature map prepares an input-dependent quantum state, state overlaps provide kernel entries, and the resulting kernel matrix is consumed by classical algorithms such as SVMs, yielding the quantum support vector machine (QSVM) workflow \cite{SchuldKilloran2019FeatureHilbert,Havlicek2019QFeatureSpace}. 

Recent studies have demonstrated quantum-kernel-based modeling across diverse application domains, from semiconductor process prediction \cite{Wang2024GaNHEMTQML} and autonomous materials science workflows \cite{Adams2026AutonomousMaterialsQKML} to trapped-ion demonstrations of QSVM-style classification and regression \cite{Suzuki2024TrappedIonQSVM}. Further extensions include tumor-metastasis classification \cite{Li2022TumorMetastasisQKernel}, high-dimensional neuroimaging pipelines for clinical staging \cite{Chen2024CompressedMediQ}, and UAV-swarm intrusion detection in network-security settings \cite{Chen2025UAVSwarmQML}.

Despite this rapid growth, QSVMs and quantum kernels still face fundamental bottlenecks in the NISQ regime. Kernel values are estimated from noisy circuits with finite measurement shots, and increased noise or effective circuit depth can amplify uncertainty and erode kernel separability. More fundamentally, multi-qubit quantum kernels can exhibit exponential concentration, where similarities across distinct inputs collapse toward near-constant values as the qubit count grows, degrading learnability and inflating the measurement budget required to resolve meaningful differences \cite{Thanasilp2024ExponentialConcentration,Thanasilp2022ExponentialConcentrationArXiv}. 

When trainable embeddings or variational kernel constructions are introduced, optimization can further suffer from barren plateaus, where gradients vanish rapidly with system size and training becomes ineffective \cite{McClean2018BarrenPlateaus}. These challenges motivate stabilization strategies that combine expressive quantum components with reliable classical structure. In particular, a quantum--classical dual kernel linearly mixes a quantum kernel with a classical kernel, using the classical component as a stabilizing anchor against concentration effects. This perspective also aligns with multiple kernel learning, which learns data-driven combinations of base kernels to improve robustness and generalization \cite{Vedaie2020QMKL,Phassadawongse2025AdaptiveDualKernel}.

A central obstacle to establishing scalable and reproducible conclusions is that large-qubit studies are constrained simultaneously by hardware noise and classical simulation limits. Statevector simulation incurs exponential memory and compute costs, making large-scale kernel-matrix construction and controlled diagnostics prohibitive beyond modest sizes. Tensor Network (TN) simulation provides a scalable alternative by representing circuits as contractible tensor graphs and exploiting contraction-path optimization and slicing for structured circuit families \cite{MarkovShi2008TN,PanZhang2022BigBatchTN,Zhang2026IterativeMPGrover}. These ideas become substantially more powerful when combined with GPU acceleration and multi-GPU parallelism \cite{Bayraktar2023cuQuantumSDK_QCE}. Building on this foundation, the present work adopts a TN-based and multi-GPU-parallel simulation pipeline to compute quantum-kernel matrix elements directly via tensor contraction, enabling near-noiseless and reproducible evaluation at scales up to hundreds of qubits and beyond \cite{Liu2024cuTNQSVM,Liu2025ValidatingLargeScaleQML,Astuti2025NQETwitterNetworkQSVC}.

\section{Method}
\subsection{Data Pre-processing}
The Fashion-MNIST~\cite{xiao2017fashion} dataset was used in all experiments for the Result section, where each $28 \times 28$ grayscale image data was flattened into a 784-dimensional feature vector. The dataset was randomly divided into 80\% for training and 20\% for testing. The raw input features were first standardized using zero-mean and unit-variance normalization to eliminate scale discrepancies across dimensions.

Principal component analysis (PCA)~\cite{abdi2010principal} was subsequently applied to decorrelate the feature space and concentrate informative variance into lower-dimensional representations. To systematically investigate the impact of input dimensionality on kernel performance, the leading $n$ principal components were retained to match different qubit numbers, with $n$ ranging from 2 to 784.

Finally, the PCA-transformed features were rescaled using min--max normalization to map all values into the interval $[0,1]$, ensuring numerical compatibility with both classical kernel evaluation and quantum feature encoding.

\subsection{Support Vector Machine for classification}
Consider a training set $\{(\mathbf{x}_i, y_i)\}_{i=1}^{m}$ with $\mathbf{x}_i \in \mathbb{R}^n$ and $y_i \in \{-1,+1\}$. The SVM is trained by solving the following dual optimization problem over Lagrange multipliers $\alpha_i \geq 0$ ~\cite{bottou2007support}:
\begin{equation}
\min_{\boldsymbol{\alpha}} \; 
\frac{1}{2} \sum_{i=1}^{m} \sum_{j=1}^{m} \alpha_i \alpha_j y_i y_j K(\mathbf{x}_i, \mathbf{x}_j)
- \sum_{i=1}^{m} \alpha_i ,
\end{equation}
subject to the constraints
\begin{equation}
\sum_{i=1}^{m} \alpha_i y_i = 0, 
\qquad 0 \leq \alpha_i \leq C,
\end{equation}
where $K(\mathbf{x}_i, \mathbf{x}_j)$ denotes a kernel function and $C$ is the regularization parameter controlling the trade-off between margin maximization and classification error.

In this work, we adopt the radial basis function (RBF) kernel as the classical kernel, defined as
\begin{equation}
    K(\mathbf{x}_i,\mathbf{x}_j)= \exp\!\left(-\gamma \lVert \mathbf{x}_i - \mathbf{x}_j \rVert^2 \right)
\end{equation}
where the hyperparameter $\gamma$ determines the sensitivity to feature distance.

This dual formulation depends solely on pairwise kernel evaluations, enabling nonlinear classification through implicit feature mappings without explicitly constructing high-dimensional representations.

After solving the dual optimization problem and obtaining the optimal Lagrange multipliers $\{\alpha_i\}_{i=1}^m$, the classification of a new sample $\mathbf{x}$ is performed using the decision function
\begin{equation}
\hat{y}(\mathbf{x}) = \mathrm{sign}\!\left( f(\mathbf{x}) \right)
\end{equation}

\begin{equation}
f(\mathbf{x}) = \sum_{i=1}^{m} \alpha_i y_i K(\mathbf{x}_i, \mathbf{x}) + b
\end{equation}

where $\hat{y}(\mathbf{x}) \in \{-1,+1\}$ denotes the predicted class label and $b$ is the bias term.

Only training samples with nonzero Lagrange multipliers $\alpha_i > 0$, referred to as support vectors, contribute to the decision function. A positive value of $f(\mathbf{x})$ corresponds to class $+1$, while a negative value corresponds to class $-1$.

Since the original SVM is formulated for binary classification, multiclass prediction is achieved by using the One-Against-One (OAO) ~\cite{debnath2004decision} strategy. For a problem with $K$ classes, $K(K-1)/2$ binary SVM classifiers were trained. During inference, each classifier casts a vote for one of the two classes, and the final class label is determined by majority voting across all pairwise classifiers.

\subsection{Quantum Support Vector Machine via Tensor Network Approach}
\begin{figure}[!t]
\centering
\includegraphics[scale=0.15]{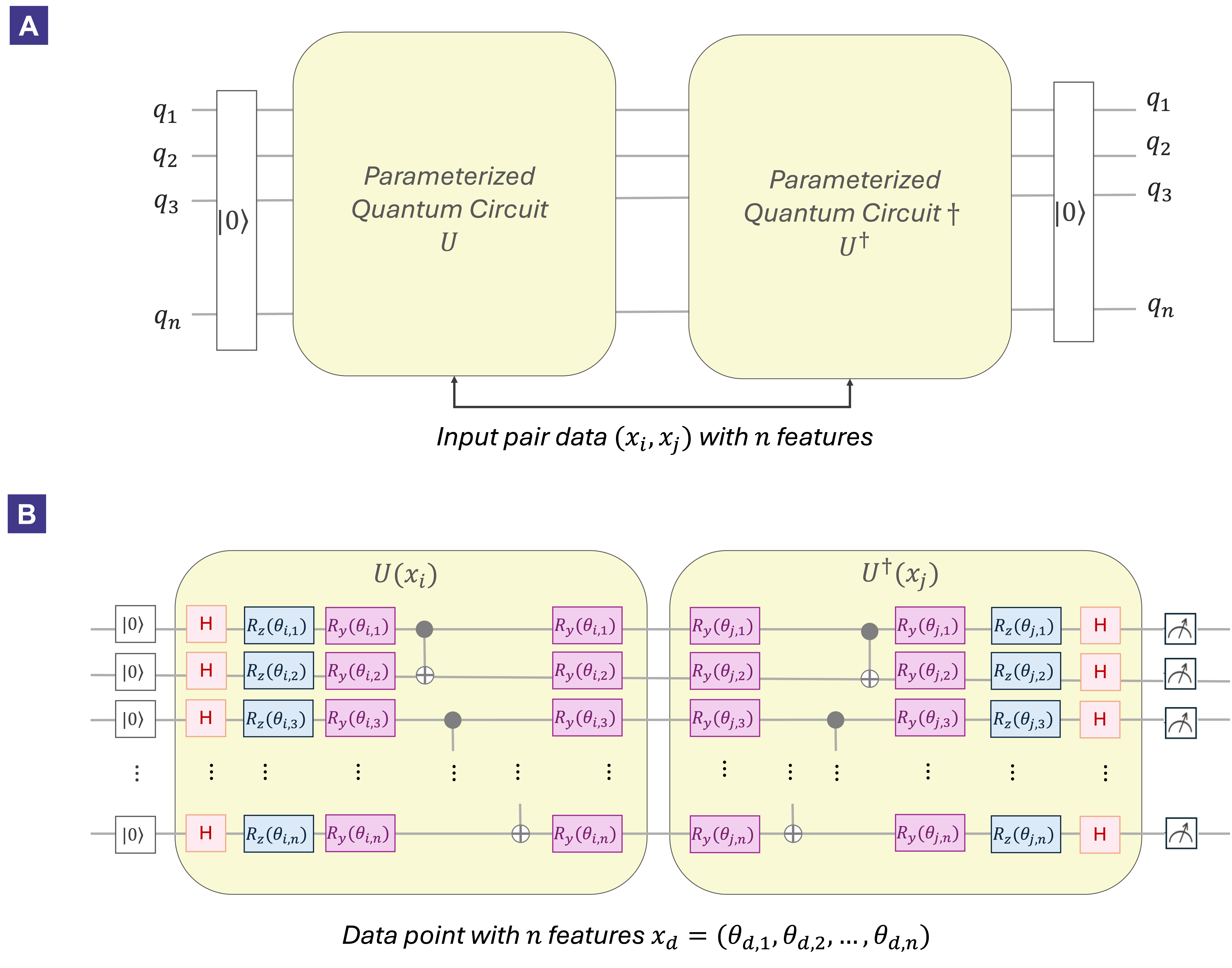}
\caption{\textbf{QSVM framework and its circuit realization.} (A) Classical data vectors are mapped to quantum states via parameterized circuits $U(x_i)$, whose overlaps are obtained through $U(x_i)U^{\dagger}(x_j)$. (B) Tensor-network–representable structure of the QSVM with the specific parameterized quantum circuit used in this work. Each data is presented as angular information of the quantum circuit.} 
\label{fig:QSVM}
\end{figure}

Figure~\ref{fig:QSVM}(A) illustrates the quantum-kernel evaluation pipeline used in a quantum support vector machine. A data-dependent feature map $U(\mathbf{x})$ embeds a classical input $\mathbf{x}$ into an $n$-qubits quantum state
\begin{equation}
\ket{\psi(\mathbf{x})} = U(\mathbf{x})\ket{0}^{\otimes n},
\label{eq:qstate}
\end{equation}
and the quantum kernel is defined by the squared overlap
\begin{equation}
K_q(\mathbf{x}_i,\mathbf{x}_j)=\left|\braket{\psi(\mathbf{x}_i)}{\psi(\mathbf{x}_j)}\right|^2 .
\label{eq:qkernel}
\end{equation}
In practice, each kernel entry is estimated using a compute--uncompute circuit that applies $U(\mathbf{x}_i)$ followed by $U^\dagger(\mathbf{x}_j)$, forming the composite unitary
\begin{equation}
W(\mathbf{x}_i,\mathbf{x}_j)=U^\dagger(\mathbf{x}_j)\,U(\mathbf{x}_i).
\label{eq:W}
\end{equation}
The all-zeros return amplitude is
\begin{equation}
a(\mathbf{x}_i,\mathbf{x}_j)=\bra{0}^{\otimes n}W(\mathbf{x}_i,\mathbf{x}_j)\ket{0}^{\otimes n},
\label{eq:amp}
\end{equation}
and the kernel value follows as
\begin{equation}
K_q(\mathbf{x}_i,\mathbf{x}_j)=|a(\mathbf{x}_i,\mathbf{x}_j)|^2 .
\label{eq:kernel_from_amp}
\end{equation}
The resulting kernel matrix can be used directly in the standard SVM dual formulation by replacing the classical kernel $K(\mathbf{x}_i,\mathbf{x}_j)$ with $K_q(\mathbf{x}_i,\mathbf{x}_j)$ \cite{SchuldKilloran2019FeatureHilbert}.

Figure~\ref{fig:QSVM}(B) depicts the featuremap structure adopted in this work. Instead of globally entangling designs, a block product state (BPS) style circuit is used, where single-qubit data-encoding layers (e.g., Hadamards and data-dependent rotations) are interleaved with locally restricted, block-structured entangling layers. This design is motivated by the observation that block-structured (locally entangled) circuits can remain expressive for image-like data while being substantially more amenable to TN simulation. 

To scale kernel evaluation beyond statevector limits, the overlap circuit in Eq.~\eqref{eq:W} is converted into an exact tensor-network by mapping each gate to a tensor and contracting the resulting tensor graph to obtain $a(\mathbf{x}_i,\mathbf{x}_j)$ in Eq.~\eqref{eq:amp}. Since constructing a kernel matrix for $m$ samples requires $O(m^2)$ pairwise evaluations, entries are computed blockwise and distributed across multiple GPUs for parallel TN contractions, reducing peak memory and increasing throughput for large-scale QSVM kernel construction\cite{Liu2025ValidatingLargeScaleQML}.

\subsection{Large-Scale Quantum--Classical Dual SVM}
\begin{figure}[!t]
\centering
\includegraphics[scale=0.2]{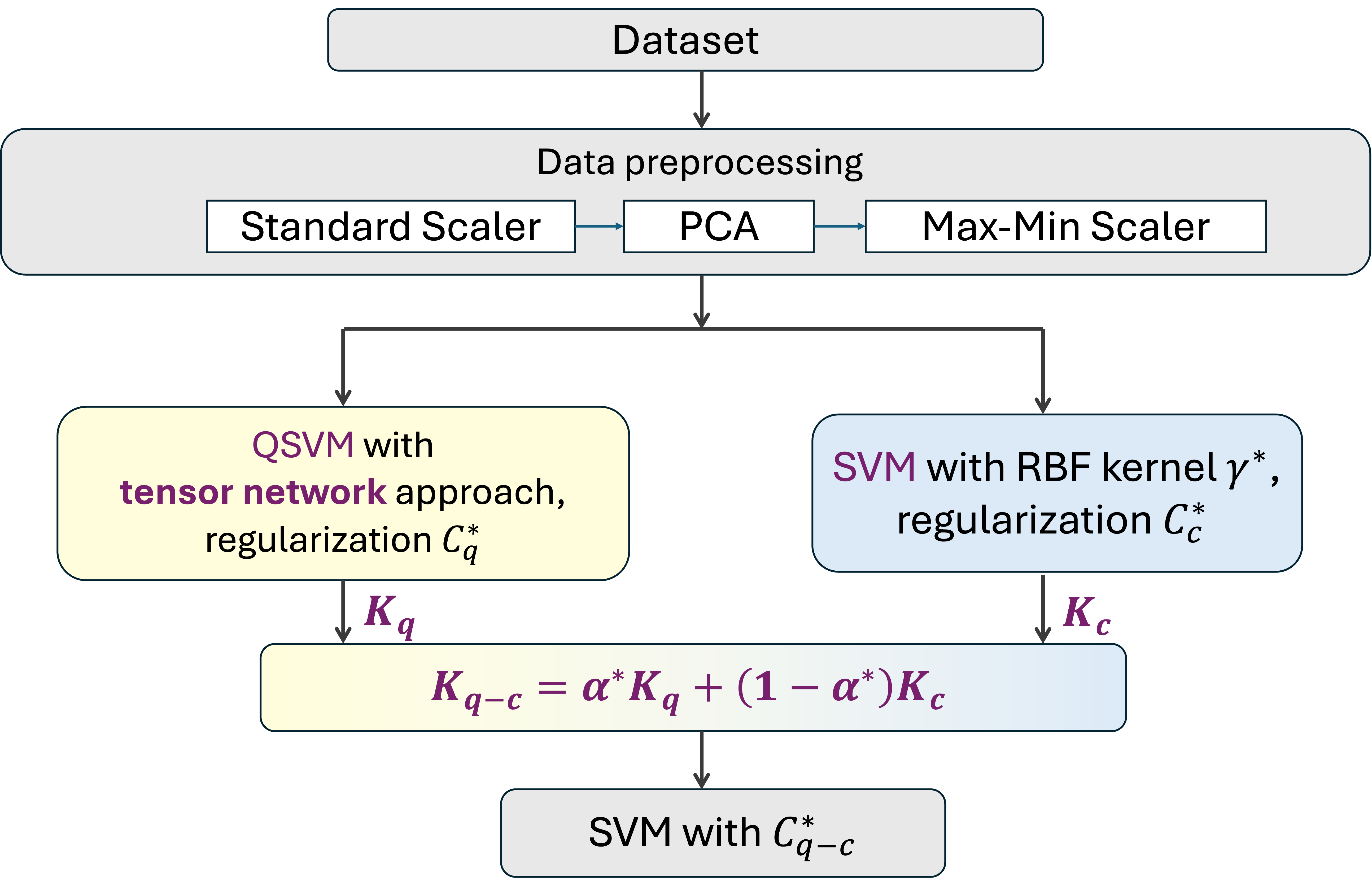}
\caption{\textbf{Framework of the quantum–classical dual-kernel SVM.} Data are first preprocessed via standard scaling, PCA, and min–max scaling, followed by parallel kernel construction using QSVM with tensor network approach and classical RBF SVM. Then the quantum $K_q$ and classical $K_c$ kernels are linearly combined to form the hybrid kernel $K_{q-c}$ which is used for final SVM classification. Note that all hyperparameter (symbols with $*$) are selected by cross-validation.
} 
\label{fig:framework}
\end{figure}

Following the adaptive quantum–classical dual kernel learning framework~\cite{Phassadawongse2025AdaptiveDualKernel}, we integrate quantum and classical kernel matrices within a unified SVM optimization scheme to leverage the complementary strengths of both feature spaces. The overall workflow is illustrated in Fig.~\ref{fig:framework}.

After data preprocessing and independent training of the classical and quantum kernel SVM models, the optimal quantum kernel matrix $K_q$ and classical kernel matrix $K_c$ are obtained with their respective tuned hyperparameters $C_q^{*}, C_{c}^{*}, \gamma^*$ by 5-fold cross-validation. Since both kernels are positive definite and share identical dimensions, they can be linearly combined to construct a hybrid kernel:

\begin{equation}
    K_{q-c}=\alpha^{*}K_{q} + (1-\alpha^{*})K_{c}
\end{equation}

Where the weighting coefficient $\alpha \in [0,1]$ controls the relative contribution between quantum and classical kernels. Note that, $\alpha$ is optimized through 5-fold cross validation to maximize classification performance. The resulting hybrid kernel 
is then used within the standard SVM dual formulation to train the final classifier (which allows its optimized $C_{q-c}^{*}$). This dual-kernel approach effectively balances the expressive capacity of quantum feature mappings with the robustness of classical kernel learning, enabling improved generalization performance.

However, constructing a full kernel matrix requires $O(m^2)$ pairwise evaluations for size of dataset $m$.
To control memory footprint and maximize GPU utilization, kernel matrices are computed in blocks. Each block $(i{:}i{+}B,\,j{:}j{+}B)$ is assigned to a GPU worker, which contracts TNs (quantum branch) or evaluates RBF values (classical branch) for the corresponding pairs, then writes the resulting block to host memory. This blockwise strategy supports multi-GPU parallelism and allows kernel construction at scales where monolithic evaluation becomes impractical. \cite{Liu2025ValidatingLargeScaleQML}.

\section{Results}
\subsection{Performance for binary classification}
We first evaluate the binary classification performance of the classical kernel, quantum kernel, and the quantum-classical dual-kernel on the Fashion-MNIST dataset across 45 pairwise class combinations. In this experiment, the number of input features/qubits is increased from low-dimensional regimes to the full 784-dimensional representation to assess model performance in large-scale settings.

\begin{figure}[!t]
\centering
\includegraphics[scale=0.35]{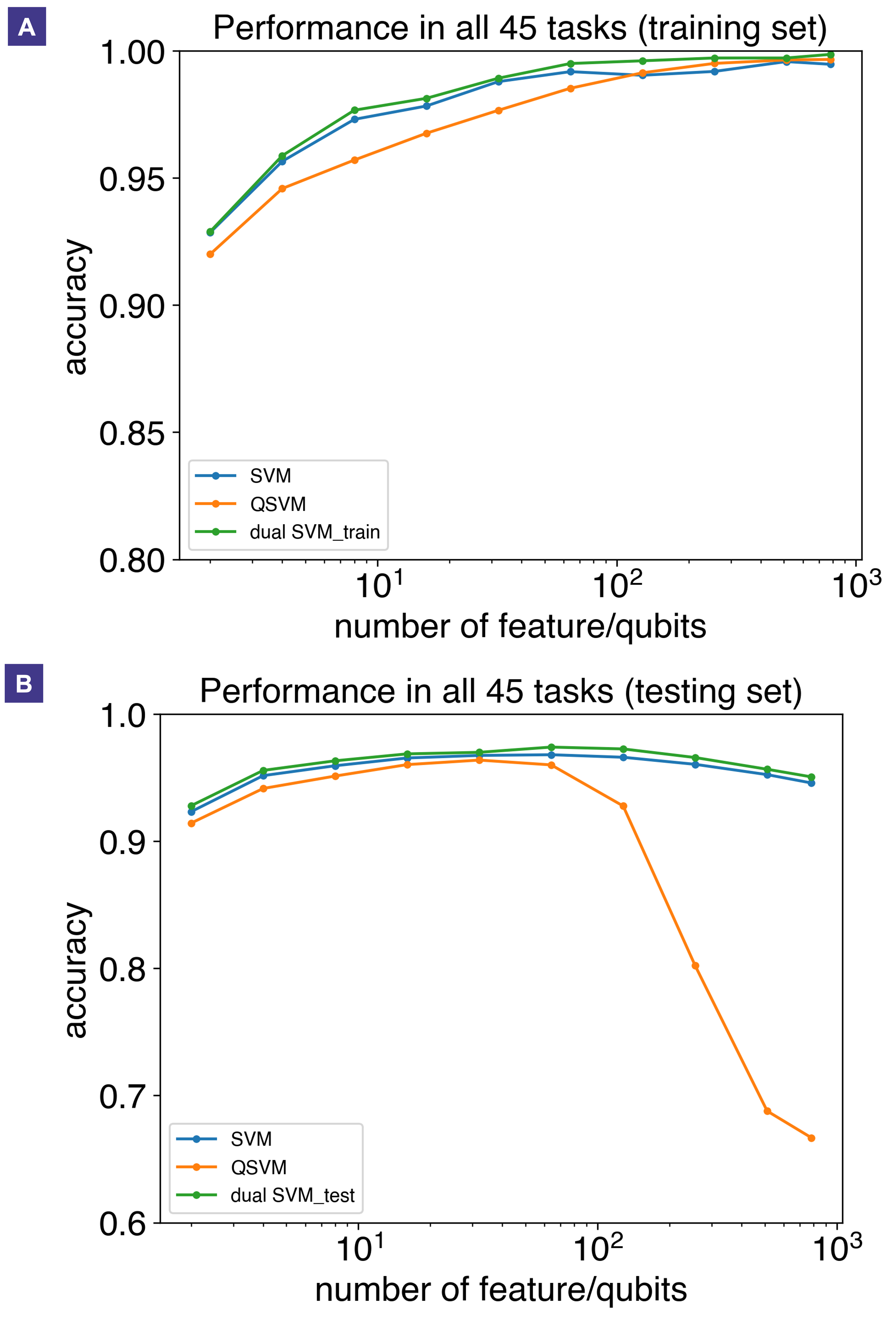}
\caption{\textbf{Accuracy as a function of the number of features/qubits for classical kernel, quantum kernel, and the quantum-classical dual-kernel across 45 binary classification tasks on the Fashion-MNIST dataset.} (A) Training-set performance shows that all models achieve increasing accuracy with higher feature dimensionality. (B)Testing-set performance initially improves with increasing model complexity but degrades at large $n$ due to overfitting, which is particularly severe for the quantum kernel model. Across all qubit regimes, the quantum-classical dual-kernel consistently achieves higher average accuracy, demonstrating its robustness as a hybrid optimization approach.
} 
\label{fig:performance}
\end{figure}

Figure ~\ref{fig:performance}(A) shows the average training accuracy across all 45 tasks as a function of the number of features. For all models, training performance improves with increasing dimensionality, reflecting the enhanced expressive capacity of the corresponding kernel feature spaces. 

On the other hand, the average testing accuracy presented in Fig.~\ref{fig:performance}(B) reveals a clear overfitting regime at large feature dimensions. While the classical kernel maintains relatively stable generalization performance, the quantum kernel exhibits an obvious accuracy degradation as the number of qubits increases (starting from $n=128$), demonstrating strong sensitivity of the model. Notably, the quantum-classical dual-kernel model preserves high test accuracy across the full range of feature/qubit numbers and consistently outperforms both individual kernel models on average.

These results indicate that increasing feature dimensionality alone does not guarantee improved generalization, particularly for highly expressive quantum kernels. By combining classical and quantum kernels within a unified dual optimization framework, the quantum-classical dual-kernel effectively balances expressivity and robustness, yielding superior and more stable performance in binary classification tasks.

\subsection{Performance for multi-class classification}
We further evaluate the multiclass classification performance of the classical kernel, quantum kernel, and the quantum-classical dual-kernel on the Fashion-MNIST dataset. We focus on the model with $n=64$ features/qubits, where all binary classifiers across different kernel models are well trained, as indicated by the training performance in Fig.~\ref{fig:performance}(A).

\begin{figure}[!t]
\centering
\includegraphics[scale=0.24]{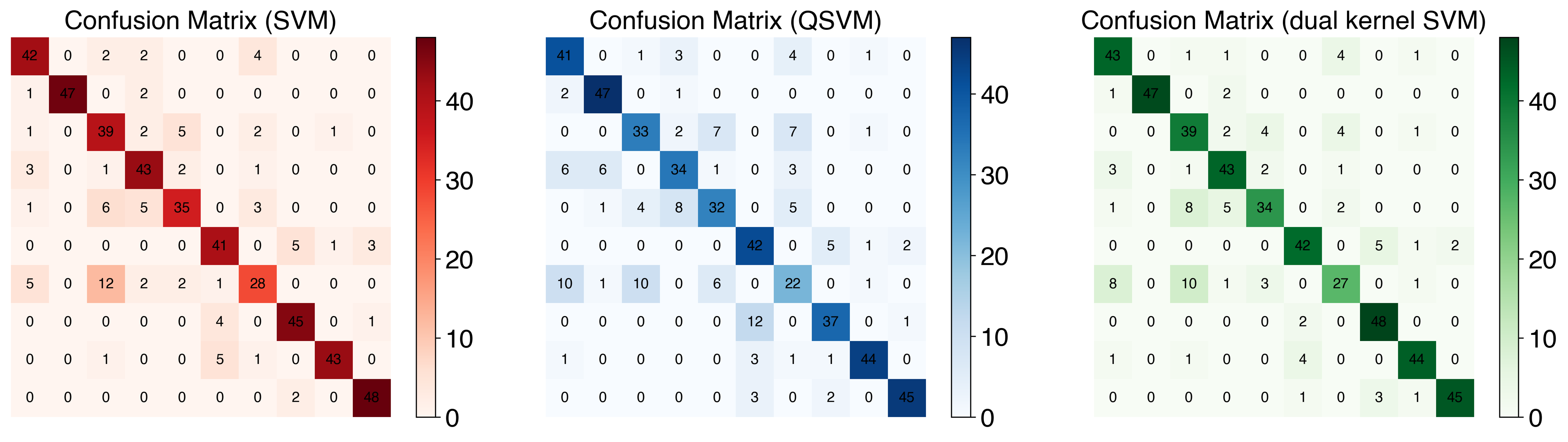}
\caption{\textbf{Confusion matrices for 10-class Fashion-MNIST classification at $n=64$ comparing SVM, QSVM, and dual-kernel SVM.} The classical and dual-kernel SVMs exhibit comparable classification performance and both outperform the pure quantum kernel model.
} 
\label{fig:confusion}
\end{figure}

Figure~\ref{fig:confusion} presents the corresponding confusion matrices for the three models. The classical kernel and the quantum-classical dual-kernel exhibit strong diagonal dominance, whereas the quantum kernel displays a larger number of off-diagonal entries, indicating substantial inter-class confusion.

Quantitative performance metrics are summarized in Table~\ref{tab:matrices}. The classical kernel and the quantum-classical dual-kernel achieve comparable overall accuracy (0.822 and 0.824, respectively), both substantially outperforming the quantum kernel, which attains an accuracy of 0.754. A similar trend is observed for the F1-score, where the quantum-classical dual-kernel slightly exceeds the classical baseline while maintaining high sensitivity. Notably, both classical and quantum-classical dual-kernel models preserve strong recall performance, whereas the quantum kernel exhibits a clear degradation across all evaluated metrics.

Overall, these results demonstrate that at moderate feature dimensionality, the quantum-classical dual-kernel approach retains the robust generalization of classical kernel learning while mitigating the performance instability of highly expressive quantum kernels in multiclass classification.

\begin{table}[!t]
\centering
\caption{Classification performance metrics for multiclass prediction at $n=64$.}
\label{tab:matrices}
\begin{tabular}{c|cccc}
\hline
\textbf{Model}  & \textbf{Accuracy} & \textbf{F1-score} & \textbf{Sensitivity} & \textbf{Recall} \\ \hline
SVM                    & 0.822             & 0.821             & \textbf{0.980}       & 0.822           \\
QSVM                   & 0.754             & 0.751             & 0.973                & 0.754           \\
Dual-kernel SVM        & \textbf{0.824}    & \textbf{0.822}    & \textbf{0.980}       & \textbf{0.824}  \\ \hline
\end{tabular}
\end{table}

\subsection{Effect of Weighting Coefficient $\alpha$}
In this section, we investigate the influence of the kernel weighting coefficient $\alpha$ on the performance of the dual-kernel SVM with a fixed feature dimension $n=64$ on binary classification task of Fashion-MNIST dataset. For each task, $\alpha$ is independently tuned to evaluate testing accuracy, and the reported results correspond to the average accuracy across all tasks.

\begin{figure}[!t]
\centering
\includegraphics[scale=0.35]{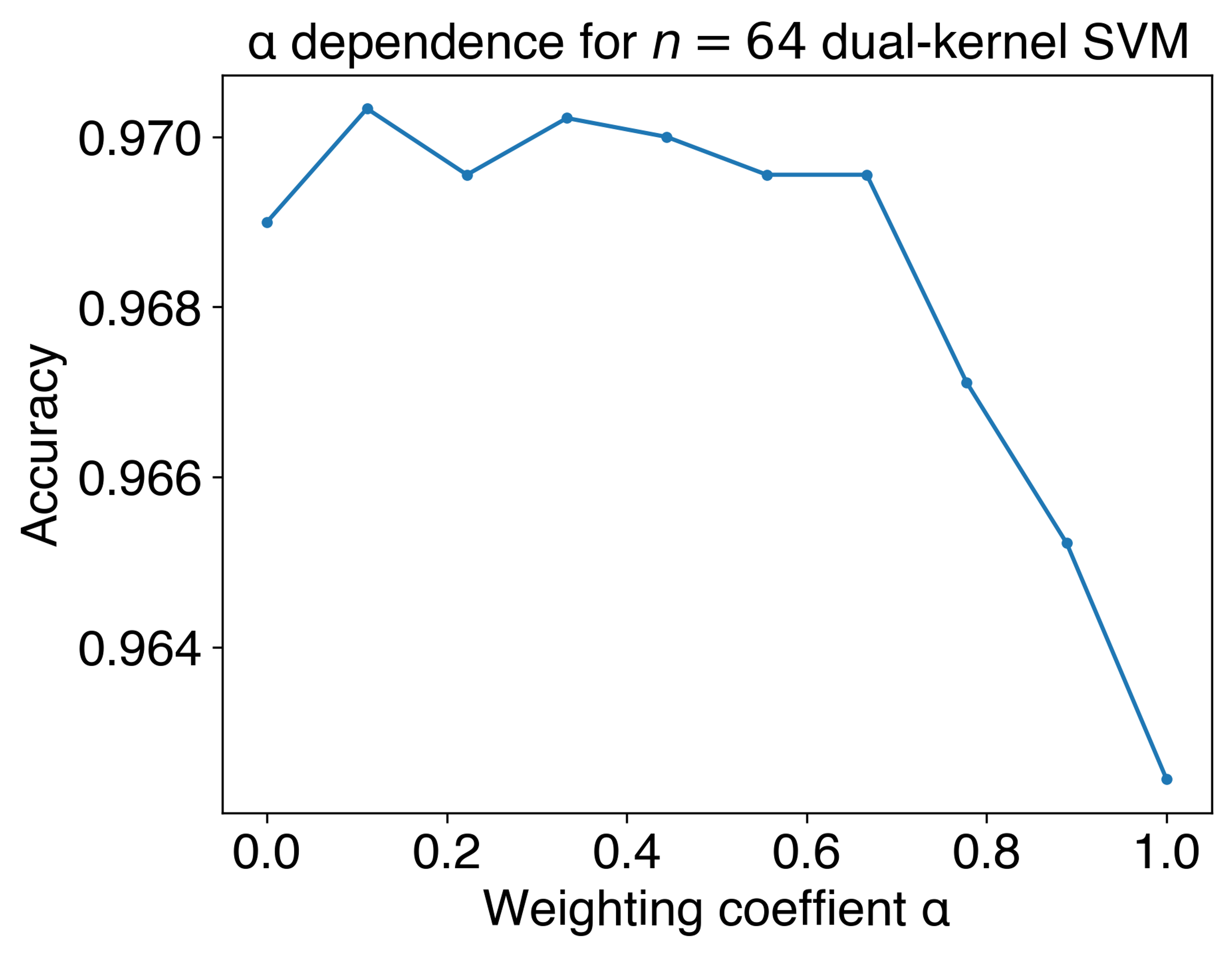}
\caption{\textbf{Average testing accuracy of the dual-kernel SVM as a function of the weighting coefficient $\alpha$ at $n=64$ over 45 binary tasks.} The accuracy curve shows optimal performance at intermediate $\alpha$ values and degradation toward the pure quantum-kernel regime.
} 
\label{fig:alpha}
\end{figure}

As shown in Fig.~\ref{fig:alpha}, the average test accuracy first increases as $\alpha$ moves away from the purely classical regime and then stays high at middle values of $\alpha$. This result shows that moderate amount of the quantum kernel enhances classification  performance (compared with the classical kernel alone). However, when $\alpha$ approaches one, where the quantum kernel dominates ($\alpha \geq 0.778$), the accuracy clearly decreases, reflecting the degraded generalization performance of quantum kernel.

These observations suggest that optimal performance arises from a balanced integration of classical and quantum kernels. In contrast, domination of the quantum kernel creates overly complex feature representations, which reduce the model’s ability to generalize. This is consistent with the overfitting behavior observed at large feature dimensions.

\section{Conclusion and discussion}


This study presents a scalable and reproducible framework for large-scale quantum–classical dual-kernel learning by combining exact tensor-network contraction with multi-GPU parallelism, enabling quantum-kernel matrix construction and SVM training up to 784 qubits. Within a unified SVM dual objective, the proposed dual kernel linearly integrates a data-dependent quantum kernel with a classical RBF kernel, providing a practical mechanism to balance expressivity and stability in large-qubit regimes.

The central conclusion is that increasing qubit dimensionality alone does not guarantee better generalization for quantum-kernel SVMs. Although all models achieve higher training accuracy as dimensionality grows, the pure quantum-kernel QSVM becomes increasingly fragile at scale, whereas the dual kernel remains consistently robust across the full range. This improvement is best explained not by added complexity, but by the classical component acting as a stabilizing anchor that preserves learnability when the quantum kernel exhibits scale-induced degradation. The weighting analysis supports this view by showing that optimal performance occurs at intermediate mixing, while performance deteriorates when the quantum component dominates.

Overall, these results position quantum–classical dual-kernel integration as a low-overhead stabilization strategy for near-term quantum kernel learning, especially when large-qubit effects and noise sensitivity limit reliability. Future work should explore adaptive and regularization-aware kernel combination beyond fixed linear mixing, and pair performance evaluation with kernel diagnostics. Extending experiments to more challenging datasets and explicit noise models will further clarify when the quantum component provides unique benefits beyond strong classical baselines.

\section*{Acknowledgment}
This work was financially supported by the National Science and Technology Council (NSTC), Taiwan. The authors would like to thank the National Center for High-performance Computing of Taiwan for providing computational and storage resources. The successful completion of this research was made possible by the academic resources and advanced research infrastructure provided by the National Center for High-Performance Computing, National Institutes of Applied Research (NIAR), Taiwan. We gratefully acknowledge their invaluable support.


\bibliographystyle{ieeetr}
\bibliography{reference} 

\vspace{12pt}
\end{document}